\documentclass[aps,prl,reprint,superscriptaddress,nofootinbib]{revtex4-2}

\usepackage{amsmath,amssymb}
\usepackage{graphicx}
\usepackage{bm}
\usepackage{booktabs}
\usepackage{microtype}

\newcommand{\Om}{\Omega}
\newcommand{\kap}{\kappa}
\newcommand{\Zinf}{Z^\infty}
\newcommand{\Edot}{\dot E}

\begin{document}

\title{Long-Lived Ringing of Near-Extremal Kerr Black Holes Resonantly Driven by Extreme-Mass-Ratio Inspirals}

\author{Wen-Biao Han}
\email{wbhan@shao.ac.cn}
\affiliation{State Key Laboratory of Radio Astronomy and Technology, Shanghai Astronomical Observatory, CAS, 80 Nandan Road, Shanghai 200030, China}
\affiliation{School of Fundamental Physics and Mathematical Sciences,
Hangzhou Institute for Advanced Study, University of Chinese Academy of
Sciences, Hangzhou 310024, China}
\affiliation{School of Astronomy and Space Science, University of Chinese
Academy of Sciences, Beijing 100049, China}

\date{\today}

\begin{abstract}
Near-extremal Kerr black holes support zero-damped modes (ZDMs), whose small
time-domain damping rates make them long-lived probes of the near-horizon
region.  We show
that bound extreme-mass-ratio inspirals (EMRIs) can resonantly drive this
response in vacuum general relativity.  Using frequency-domain Teukolsky
amplitudes for eccentric-inclined Kerr geodesics, we identify a
source-supported orbital harmonic whose real frequency falls within one pole
half-width of the fundamental gravitational ZDM.  In the complex response, the
pole contribution is enhanced by this small half-width, while complex-response
tomography recovers the independently computed Kerr pole from real-frequency
orbital data.  After subtracting the smooth non-pole component, the residual
exhibits the phase winding of a coherent simple pole, with a pole contribution
comparable to the smooth non-pole part of the EMRI-sourced Teukolsky amplitude.
The driven branch also lies in the superradiant regime and carries negative
horizon flux.  These results establish a pole-resolved, resonantly driven ZDM
response by EMRIs and make the recovered pole half-width a route to measuring
the horizon surface gravity.
\end{abstract}

\maketitle

\begin{figure*}[t]
 \includegraphics[width=\textwidth]{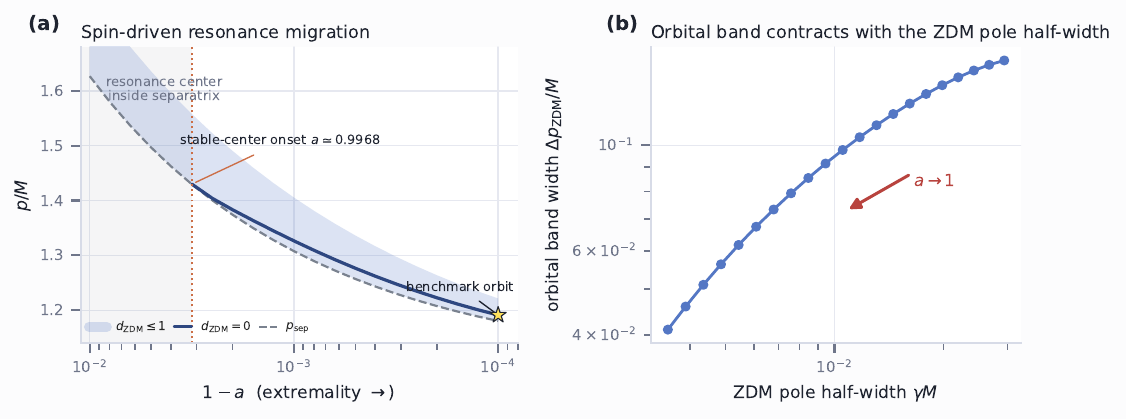}
 \caption{\label{fig:branch}
 \textbf{Spin-driven migration and contraction of the ZDM resonance channel.}
 Representative benchmark with $e=0.1$, $\iota=22^\circ$, and the
 $(m,n,k)=(2,2,1)$ harmonic.
 (a) Stable $d_{\rm ZDM}\leq1$ band versus $1-a$.  The solid curve satisfies
 $d_{\rm ZDM}=0$, shading denotes $d_{\rm ZDM}\leq1$, and the dashed curve is
 the inclined separatrix, where
 $d_{\rm ZDM}=|\omega_{221}-\operatorname{Re}\omega_{\rm ZDM}|/\gamma$ and
 $\gamma=-\operatorname{Im}\omega_{\rm ZDM}$.  In the gray region the exact
 resonance center lies inside the separatrix, although stable orbits still
 enter the $d_{\rm ZDM}\leq1$ band.  The star marks the $a=0.9999$ benchmark orbit
 used for the tomography and flux diagnostics.
 (b) Orbital band width
 $\Delta p_{\rm ZDM}\equiv p_{\rm out}-p_{\rm in}$, the stable connected
 $d_{\rm ZDM}\leq1$ interval, versus the independent ZDM pole half-width.}
\end{figure*}

\textit{Introduction.---}
Extreme-mass-ratio inspirals (EMRIs) are among the cleanest probes of
strong-field Kerr dynamics.  A compact object orbiting a massive black hole
radiates over $10^4$--$10^5$ cycles, and its waveform encodes the three
fundamental frequencies of generic bound Kerr geodesics
\cite{Schmidt2002,DrascoHughes2004,FujitaHikida2009},
\begin{equation}
 \omega_{mnk}=m\Om_\phi+n\Om_r+k\Om_\theta ,
 \label{eq:harmonic}
\end{equation}
This discrete spectrum
underlies black-hole mapping and precision tests of general relativity with
space-based gravitational-wave detectors
\cite{Ryan1995,BarackCutler2004,Gair2013,AmaroSeoane2017,Babak2017}.  Most EMRI
studies use these frequencies as probes of the background geometry and of the
orbital radiation reaction.  Here we ask a different question: can an EMRI
harmonic act as a coherent external driver of a long-lived black-hole mode?

The relevant modes are the zero-damped quasinormal modes (ZDMs) of
near-extremal Kerr black holes.  Kerr perturbations are governed by the
Teukolsky equation \cite{Teukolsky1973}, and the associated quasinormal
spectrum has been studied extensively since the classic work of Detweiler and
Leaver \cite{Detweiler1980,Leaver1985,Berti2009}.  Near extremality, the
spectrum separates into damped modes and ZDMs, whose imaginary parts vanish in
the extremal limit \cite{Yang2013}.  For a ZDM,
\begin{equation}
 \omega_{\ell m N}^{\rm ZDM}
 =m\Om_H+\kap\bigl(\varpi_{\ell m N}
 -i\Gamma_{\ell m N}\bigr)+O(\kap^2),
 \label{eq:zdm-expansion}
\end{equation}
where $N$ is the overtone index and
\begin{align}
 \varpi_{\ell m N}
 &\equiv\lim_{\kap\to0}
 \frac{\operatorname{Re}\omega_{\ell m N}^{\rm ZDM}-m\Om_H}{\kap},
 \nonumber\\
 \Gamma_{\ell m N}
 &\equiv\lim_{\kap\to0}
 \frac{-\operatorname{Im}\omega_{\ell m N}^{\rm ZDM}}{\kap}>0 .
 \label{eq:zdm-coefficients}
\end{align}
Thus $\varpi_{\ell m N}$ is the dimensionless real-frequency offset from the
superradiant bound, while $\Gamma_{\ell m N}>0$ is the dimensionless damping
coefficient.  The physical damping rate, equivalently the pole half-width in a
real-frequency response, is therefore
\[
 \gamma\equiv-\operatorname{Im}\omega_{\ell mN}^{\rm ZDM}
 =\kap\Gamma_{\ell mN}+O(\kap^2).
\]
Here $\kap$ is the surface gravity of the Kerr horizon generator
$\chi=\partial_t+\Om_H\partial_\phi$; its explicit Kerr value is given below.
Surface gravity is the intrinsic redshift and acceleration scale of a
stationary horizon.  It enters the black-hole first law, sets the Hawking
temperature $T_H=\kap/(2\pi)$ in units with $\hbar=k_B=1$, and vanishes in the
extremal limit \cite{Bardeen1973,Hawking1975}.  Equations
(\ref{eq:zdm-expansion}) and (\ref{eq:zdm-coefficients}) show that the same
horizon scale controls the imaginary part of the ZDM frequency.  Thus the ZDM
sector is a natural candidate for horizon spectroscopy: measuring the pole
half-width is, in principle, measuring a quantity controlled directly by the
near-horizon geometry.

There are two reasons why an EMRI may access this sector more effectively than
an ordinary perturbation.  First, the source spectrum is not broadband noise
but a set of long-lived, phase-coherent orbital harmonics.  Second, near the
horizon of a rapidly spinning black hole, three effects occur simultaneously:
orbital harmonics approach the superradiant frequency scale $m\Om_H$, source
amplitudes can be enhanced near the separatrix, and the ZDM pole half-width
shrinks with $\kap$.  These ingredients suggest a driven near-horizon resonance, but
they also create an ambiguity.  A large response near the separatrix need not
be a QNM pole; it could be ordinary Teukolsky scattering, a smooth source
enhancement, or superradiant horizon kinematics.

Previous work found Kerr QNM ``wiggles'' in compact-object fields and
self-forces, while direct tuning to QNM frequencies produced only small changes
in the parameter ranges explored \cite{Thornburg2020}.  Superradiant horizon
fluxes and tidal acceleration in EMRIs have also been studied
\cite{PressTeukolsky1972,TeukolskyPress1974,Hughes2000}.  These
results motivate, but do not answer, the pole-specific question addressed
here: whether the real-frequency EMRI response contains a recoverable and
dynamically significant ZDM pole contribution.

To address this question, this Letter uses frequency-domain Teukolsky
amplitudes for eccentric-inclined Kerr geodesics.  The analysis has three
steps.  We first locate source-supported orbital regions whose harmonics enter
one ZDM pole half-width.  We then separate a shared simple-pole contribution
from smooth non-pole terms using complex-response tomography.  Finally, we
evaluate the pole fraction and the associated horizon flux along the same
branch.  This keeps pole identification separate from ordinary source
enhancement, smooth scattering, and superradiant horizon kinematics.

\begin{figure*}[t]
 \includegraphics[width=\textwidth]{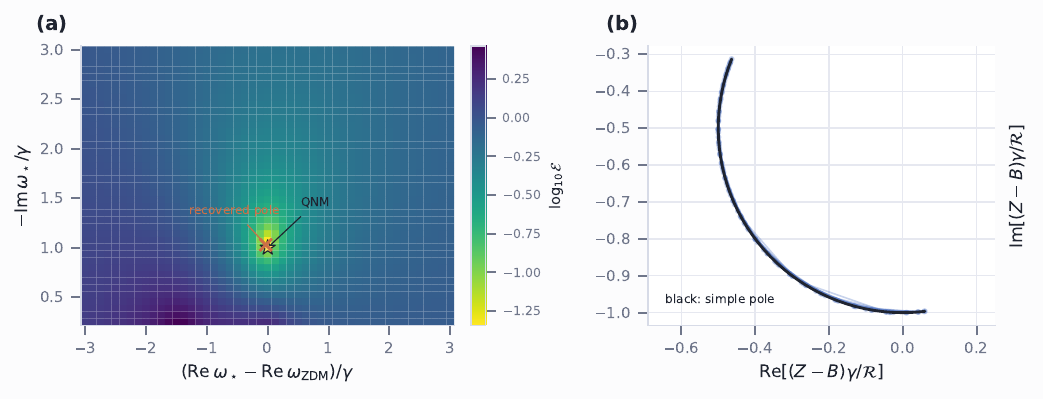}
 \caption{\label{fig:tomography}
 \textbf{Complex-response tomography selects the independent ZDM pole.}
 Shared-pole test across multiple inclination cuts, each with its own smooth
 non-pole component and residue.
 (a) Joint normalized prediction error in the complex trial-pole plane; the
 star is the independent QNM and the cross is the recovered pole.
 (b) Residue-normalized, non-pole-subtracted trajectories for all inclination
 cuts compared with a simple-pole response (black).}
\end{figure*}

\textit{Orbital access to the ZDM pole half-width.---}
We set $G=c=M=1$ and use
\begin{equation}
 r_+=1+\sqrt{1-a^2},\quad
 \Om_H=\frac{a}{2r_+},\quad
 \kap=\frac{\sqrt{1-a^2}}{2r_+}.
 \label{eq:horizon}
\end{equation}
The calculation combines Kerr geodesic frequencies computed with
\texttt{kerrgeopy} \cite{Schmidt2002,FujitaHikida2009,DrascoHughes2004},
independent Kerr QNM frequencies computed with \texttt{qnm}
\cite{qnmPackage}, and frequency-domain Teukolsky amplitudes and fluxes
computed with \texttt{pybhpt} \cite{Teukolsky1973,pybhptPackage}.  As a
representative benchmark, we use $e=0.1$, the orbital
harmonic $(m,n,k)=(2,2,1)$, and the gravitational mode
$(s,\ell,m,N)=(-2,2,2,0)$.  The near-horizon condition is
$\omega_{221}\simeq m\Om_H+\kap\varpi_{220}$, with the common azimuthal number
$m=2$ for the orbital harmonic and the ZDM.  This is the fundamental low-lying
co-rotating gravitational ZDM sector; the low-order $(n,k)$ prescreen and
neighboring co-rotating sectors are summarized in the Supplemental Material.

For each spin we compute the independent ZDM on a continuous Kerr spin
sequence and define the pole-half-width detuning
\begin{equation}
 d_{\rm ZDM}\equiv
 \frac{|\omega_{221}-\operatorname{Re}\omega_{\rm ZDM}|}
 {-\operatorname{Im}\omega_{\rm ZDM}},
 \label{eq:halfwidth}
\end{equation}
with $\omega_{\rm ZDM}=\omega_{220}^{\rm ZDM}(a)$ and
$\gamma\equiv-\operatorname{Im}\omega_{\rm ZDM}$.  The quantity
$d_{\rm ZDM}$ is a detuning normalized by the pole half-width, not an arbitrary
orbital tolerance.  Near the ZDM pole the sourced response contains
\begin{align*}
 Z_{\rm pole}(\omega)
 &=\frac{{\cal R}}{\omega-\omega_{\rm ZDM}}
 =\frac{{\cal R}}{\Delta\omega+i\gamma},\\
 \Delta\omega&\equiv
 \omega_{221}-\operatorname{Re}\omega_{\rm ZDM},
\end{align*}
so that
\[
 |Z_{\rm pole}|=
 \frac{|{\cal R}|/\gamma}{\sqrt{1+d_{\rm ZDM}^2}} .
\]
Thus $d_{\rm ZDM}=0$ is the real-frequency resonance center, while
$d_{\rm ZDM}\leq1$ denotes the region within one pole half-width, where the
driven pole amplitude remains within a factor $1/\sqrt{2}$ of its resonant value.  The
zero-damped nature of the mode enters through the small half-width
$\gamma\propto\kap$: as $a\to1$, the resonance width narrows in frequency while
the peak pole response scales as $1/\gamma$, provided the source
projection ${\cal R}$ remains nonzero and finite.

Figure~\ref{fig:branch} summarizes the orbital origin and near-extremal
scaling of this resonance channel.  The real-frequency center
$d_{\rm ZDM}=0$ migrates inward as $a\to1$ and remains close to the inclined
separatrix once a stable exact crossing appears.  The shaded region is the set
of stable orbits whose source harmonic lies within one ZDM pole half-width,
$d_{\rm ZDM}\leq1$.  Here the inclined separatrix
$p_{\rm sep}(e,\iota,a)$ denotes the last stable
orbit boundary between stable bound eccentric-inclined geodesics and plunging
motion; stable orbits require $p>p_{\rm sep}$.  Near an unclipped stable
crossing at $p=p_c$,
\[
 \omega_{221}(p)\simeq \operatorname{Re}\omega_{\rm ZDM}
 +\left.\partial_p\omega_{221}\right|_{p_c}(p-p_c),
\]
so the corresponding orbital interval obeys
\[
 \Delta p_{\rm ZDM}\sim
 \frac{2\gamma}{|\partial_p\omega_{221}|_{p_c}}.
\]
When the inner side of this interval lies inside the separatrix, the stable
band is clipped at $p_{\rm sep}$.  Thus the orbital access window should
contract as the ZDM pole half-width narrows, modulo the local orbital-frequency
gradient and separatrix clipping.

Figure~\ref{fig:branch}(b) verifies this expectation directly.  Along the
fixed-orbit family the measured stable band width
$\Delta p_{\rm ZDM}\equiv p_{\rm out}-p_{\rm in}$, where
$[p_{\rm in},p_{\rm out}]$ is the connected stable interval satisfying
$d_{\rm ZDM}\leq1$, contracts with the independently computed pole half-width.
Since the same ZDM branch has $\gamma\propto\kap$, the resonance channel is
tied to the surface-gravity scale of the near-horizon pole, rather than to a
freely chosen orbital window.  Descriptive fit coefficients are given in the
Supplemental Material.

\textit{Complex-pole tomography.---}
On each fixed-inclination cut, we write the corrected infinity amplitude as a
smooth non-pole component plus a candidate pole term,
\begin{equation}
 \Zinf_i(\omega)=B_i(\omega)+Z_{i,\rm pole}^\infty(\omega),\qquad
 Z_{i,\rm pole}^\infty(\omega)=\frac{{\cal R}_i}{\omega-\omega_\star},
\label{eq:polemodel}
\end{equation}
where $B_i(\omega)\equiv Z^\infty_{i,\rm np}$ denotes the smooth non-pole
component of the EMRI-sourced Teukolsky amplitude on inclination cut $i$.
It absorbs the direct and non-resonant scattering response of the source, as
well as other slowly varying non-pole contributions in the local frequency
window.

To implement this separation, we approximate the smooth term locally by
\begin{equation}
 B_i(\omega)=\sum_{j=0}^{2}c_{ij}x^j ,
 \label{eq:Bi}
\end{equation}
where $x$ is a centered and scaled real frequency.  This quadratic form is a
local nuisance expansion, not a global model of the EMRI response.  Each
inclination cut has its own $B_i$ and ${\cal R}_i$, while the complex pole
frequency $\omega_\star$ is shared by all cuts.  An equal-complexity cubic
polynomial supplies the no-pole control.  We compare
the models by leave-one-out complex prediction error and define the joint
score
\begin{equation}
 {\cal E}(\omega_\star)=
 \exp\!\left[
  \frac{1}{N_\iota}\sum_i
  \ln\frac{\epsilon_{{\rm pole},i}(\omega_\star)}
           {\epsilon_{{\rm cubic},i}}
 \right].
 \label{eq:score}
\end{equation}
This normalization prevents high-amplitude cuts from determining the pole.
The purpose of this tomography is to decide whether the frequency access in
Fig.~\ref{fig:branch} corresponds to an actual pole of the sourced EMRI
response, rather than to a smooth source enhancement or a single fine-tuned
orbital cut.  The test is deliberately overconstrained: each inclination cut
has its own smooth non-pole component and residue, but all cuts must share one
complex pole.

Figure~\ref{fig:tomography}(a) gives the existence and localization test.  The
joint prediction error has a compact minimum at the independently computed
ZDM, and allowing the pole to move gives
\begin{equation}
 \frac{\operatorname{Re}(\omega_\star-\omega_{\rm ZDM})}{\gamma}
 =-0.0313,\qquad
 \frac{-\operatorname{Im}\omega_\star}{\gamma}=1.029 .
\end{equation}
Thus the pole can be recovered from real-frequency EMRI data without imposing
the QNM frequency as an input.  The independent pole gives
${\cal E}=0.0529$; deliberate frequency shifts, halving or doubling the pole
half-width, and the upper-half-plane conjugate yield much larger errors, while the minimum is
unchanged when any one of the 12 inclination cuts is removed.  The
one-dimensional coordinate slices, shown in the Supplemental Material, verify
that both the real part and the pole half-width are selected.  After subtracting the
fitted smooth component and normalizing by the residue, all cuts follow the phase
winding of a simple pole, $1/(\omega-\omega_{\rm ZDM})$
[Fig.~\ref{fig:tomography}(b)].  These signatures establish that the EMRI
response contains a coherent, accessible ZDM pole contribution; its
quantitative strength relative to the smooth non-pole component is measured
next.

\begin{figure}[t]
 \includegraphics[width=\columnwidth]{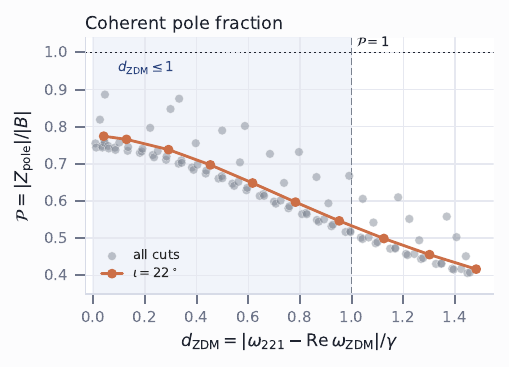}
 \caption{\label{fig:pole_fraction}
 \textbf{Coherent pole fraction.}
 Pole-to-non-pole ratio for the $a=0.9999$, $e=0.1$ tomography sample,
 plotted directly against the detuning normalized by the pole half-width.  Gray
 points show all inclination cuts; orange points highlight the benchmark
 $\iota=22^\circ$ cut.  The shaded region is the $d_{\rm ZDM}\leq1$ interval,
 and the dotted horizontal line marks pole--non-pole parity.}
\end{figure}

\textit{Pole fraction and flux.---}
To quantify dynamical importance within Eq.~(\ref{eq:polemodel}), we define
\begin{equation}
 {\cal P}=\frac{|Z_{\rm pole}|}{|B|}
 =\frac{|{\cal R}/(\omega-\omega_{\rm ZDM})|}{|B(\omega)|}.
 \label{eq:fraction}
\end{equation}
Figure~\ref{fig:pole_fraction} shows that the recovered pole is not merely
identifiable, but dynamically relevant.  Across all tomography cuts,
${\cal P}$ increases systematically as $d_{\rm ZDM}$ decreases, demonstrating
that the enhanced coherent component is tied to the ZDM pole half-width rather
than to a monotonic inward source enhancement.  The full sample has median
${\cal P}=0.632$ and maximum $0.887$: the pole remains comparable to, but
does not dominate, the smooth non-pole response.  This is consistent with the
absence of a sharp Lorentzian feature in the total amplitude.

The net energy flux gives a related energy-budget diagnostic.  We sum over
spheroidal and orbital harmonics and evaluate
\begin{equation}
 \eta_{\rm tot}=
 \frac{|\Edot_H^{\rm tot}|}{\Edot_\infty^{\rm tot}} .
 \label{eq:eta}
\end{equation}
The horizon flux remains negative throughout the common-truncation scan and
varies smoothly through the real-frequency crossing.  A higher-order
calculation with
$\ell\leq10$, $|k|\leq6$, and $|n|\leq30$ gives
$\eta_{\rm tot}=0.184$ at the high-order anchor (see Supplemental Material).
Thus the ZDM channel produces an ${\cal O}(10\%)$ superradiant correction to
the EMRI energy budget, while its smooth variation explains why the pole is
most cleanly resolved in the complex amplitude rather than in a flux peak.

As a consistency check on the driven response, compare the free decay time with
the EMRI frequency-sweep time.  In the phase frame of the orbital harmonic, the
fitted pole term obeys the reduced driven-mode equation
\begin{equation}
 \dot A+(\gamma-i\delta)A=-i{\cal R},\qquad
 \delta=\omega_{221}-\operatorname{Re}\omega_{\rm ZDM}.
 \label{eq:driven}
\end{equation}
Besides $\tau_{\rm ZDM}=\gamma^{-1}$, define the local time to sweep by one
pole half-width and its ratio to the damping time,
\begin{equation}
 \tau_{\rm sweep}=\frac{\gamma}{|\dot\delta|},\qquad
 \frac{\tau_{\rm sweep}}{\tau_{\rm ZDM}}
 =\frac{\gamma^2}{|\dot\delta|}.
 \label{eq:timescales}
\end{equation}
For $\mu/M=10^{-5}$, a reduced inspiral driven by the candidate harmonic gives
$|\dot\delta|=4.69\times10^{-10}M^{-2}$ at the crossing.  Thus
\begin{align}
 \tau_{\rm ZDM}&=288.3M,\\
 \tau_{\rm sweep}&=7.40\times10^6M
 =2.57\times10^4\tau_{\rm ZDM}.
 \label{eq:timescale-numbers}
\end{align}
For a $10^6M_\odot$ primary these are $23.7$ minutes and $1.15$ yr,
respectively.  Since $|\dot\delta|\propto\mu/M$ in the adiabatic limit, the
hierarchy grows inversely with mass ratio.  The numerical envelope follows
its instantaneous forced solution to within $4.9\times10^{-4}$ on the sampled
track.  The large separation in Eq.~(\ref{eq:timescale-numbers}) therefore
supports a quasi-steady, coherently driven ZDM rather than a short transient or
an isolated waveform spike.  A quantitative waveform or dephasing prediction
would require a full multimode inspiral through the narrow ZDM region.

\textit{Discussion.---}
The calculation establishes a specific route by which an EMRI can drive a
long-lived near-horizon ZDM response.  A discrete orbital harmonic enters one
pole half-width of the fundamental co-rotating ZDM; the corresponding access
window contracts with the independently computed pole half-width;
complex-response tomography then recovers the same pole from real-frequency
source data; and after subtracting the smooth non-pole response, the residual
follows the phase winding of a simple pole with ${\cal P}=O(1)$.  Because
$\gamma\propto\kap$, the recovered pole half-width connects this driven ZDM
response to the horizon surface gravity.  The physical point is that an EMRI
can act not only as a long-duration mapper of Kerr orbital frequencies, but
also as a coherent external driver of the near-horizon mode spectrum.  The
result is therefore not simply a large Teukolsky amplitude near the
separatrix, nor a kinematic superradiant flux effect, but a pole-resolved
response of the near-extremal horizon spectrum.

Superradiant braking is a related energy-budget consequence, not the
pole-identification criterion.  It does not follow that the flux must peak: the
complex interference between pole and the smooth non-pole component is lost in
$|\Zinf|^2$, while the horizon flux also contains smooth superradiant factors.
The timescale estimate above indicates that the candidate is driven
quasi-steadily across the pole half-width, but turning this pole-resolved
response into an observable waveform prediction requires projecting the source
onto the QNM with a properly normalized Kerr Green-function residue
\cite{Green2023} and evolving a full multimode inspiral through the
near-resonant region.  With such a calibrated waveform model, recovering the
pole half-width from an observed EMRI would provide a dynamical measurement of
the horizon surface gravity, rather than only a consistency test of the Kerr
orbital frequency map.

\textit{Acknowledgments.---}
This work was supported by the National Science and Technology Major Project of China (No. 2024ZD1100601), the National Key R\&D Program of China (No. 2021YFC2203002) and the NSFC (National Natural Science Foundation of
China) No. 12473075.

\bibliography{paper_draft}

\clearpage
\setcounter{figure}{0}
\setcounter{table}{0}
\setcounter{equation}{0}
\renewcommand{\thefigure}{S\arabic{figure}}
\renewcommand{\thetable}{S\Roman{table}}
\renewcommand{\theequation}{S\arabic{equation}}
\section*{Supplemental Material}

\section{Branch selection and near-horizon continuation}

The low-order prescreen tests whether a sourced EMRI harmonic can enter one
ZDM pole half-width in the stable region outside the inclined separatrix.  The
scan covers positive and negative radial and polar harmonics $(n,k)$, evaluates
the scaled detuning
\begin{equation}
 d_{\rm ZDM}=
 \frac{|\omega_{mnk}-\operatorname{Re}\omega_{\rm ZDM}|}{\gamma},
\end{equation}
and ranks source-supported samples by
$S_\infty=|\Zinf|^2/(1+d_{\rm ZDM}^2)$.  The selected
$(m,n,k)=(2,2,1)$ branch is then followed along a continuous near-extremal
spin sequence.  At each spin we solve for the outermost stable root of
$\omega_{221}=\operatorname{Re}\omega_{\rm ZDM}$; when the inner
$d_{\rm ZDM}=1$ boundary is unstable, the stable portion of the band is
truncated at the inclined separatrix.

\begin{table*}[t]
 \caption{\label{tab:nk_prescreen}
 Low-order sourced prescreen at $a=0.99$, $e=0.1$, $m=2$, for the
 fundamental gravitational sector $(s,\ell,m,N)=(-2,2,2,0)$.
 Here $d_{\min}$ is the minimum detuning normalized by the pole half-width on
 the coarse $p$--inclination grid, $d_T$ is the corresponding detuning at the
 Teukolsky source-supported sample, and
 $S_T\equiv |\Zinf|^2/(1+d_T^2)$.  The column
 $\dot E_H<0$ refers to the same Teukolsky sample.  The daggered
 ${\cal P}_{\max}$ is the maximum pole fraction from the full
 $a=0.9999$ tomography of the selected branch; entries marked by dashes were
 not subjected to pole/non-pole tomography.}
 \begin{ruledtabular}
 \begin{tabular}{ccccccc}
 $(n,k)$ & $d_{\min}$ & $d_T$ & ${\cal P}_{\max}$ &
 $|\Zinf|^2$ & $S_T$ & $\dot E_H<0$ \\
 \hline
 $(1,1)$ & $8.9\times10^{-2}$ & $0.619$ & -- & $1.83$ & $1.32$ & yes \\
 $(2,1)$ & $3.7\times10^{-4}$ & $0.299$ & $0.887^\dagger$ &
 $1.34$ & $1.23$ & yes \\
 $(0,1)$ & $5.8\times10^{-1}$ & $0.579$ & -- & $2.31\times10^{-1}$ &
 $1.73\times10^{-1}$ & yes \\
 $(2,2)$ & $1.0\times10^{-3}$ & $0.251$ & -- & $9.98\times10^{-2}$ &
 $9.39\times10^{-2}$ & yes \\
 $(0,2)$ & $5.0\times10^{-3}$ & $0.152$ & -- & $3.36\times10^{-2}$ &
 $3.28\times10^{-2}$ & no \\
 $(-2,2)$ & $2.0\times10^{-3}$ & $0.502$ & -- & $2.04\times10^{-2}$ &
 $1.63\times10^{-2}$ & yes \\
 $(1,2)$ & $1.8\times10^{-2}$ & $0.235$ & -- & $1.18\times10^{-2}$ &
 $1.12\times10^{-2}$ & yes \\
 $(-1,2)$ & $9.2\times10^{-3}$ & $0.113$ & -- & $1.54\times10^{-4}$ &
 $1.52\times10^{-4}$ & no
 \end{tabular}
 \end{ruledtabular}
\end{table*}

Table~\ref{tab:nk_prescreen} shows that the selected $(n,k)=(2,1)$ branch is
simultaneously close to the fundamental ZDM pole, source supported, and in the
superradiant regime.  Neighboring co-rotating sectors for the same orbital
harmonic are less competitive: the fundamental $(\ell,m,N)=(2,2,0)$ sector
lies within the pole half-width, whereas the nearby $\ell=3$ and $\ell=4$
sectors are several pole half-widths away at the source-supported point and
have weaker source support.  The table is a branch-selection diagnostic rather
than a uniqueness claim.

\begin{figure*}[t]
 \includegraphics[width=\textwidth]{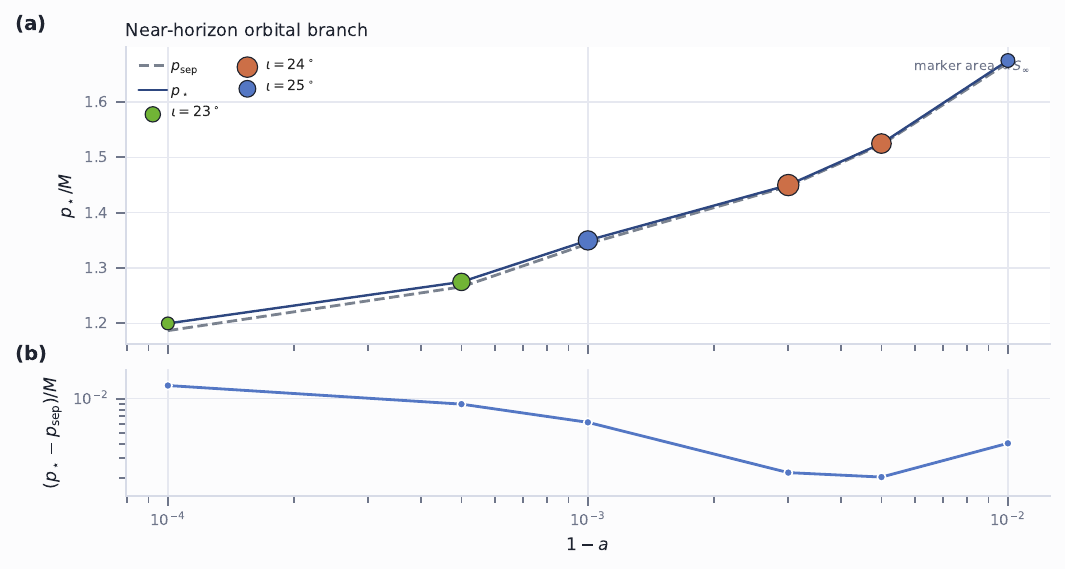}
 \caption{\label{fig:supp_spin_branch}
 Source-supported spin branch.  Top: selected semi-latus rectum and inclined
 separatrix, with marker area proportional to $S_\infty$ and color denoting
 inclination.  Bottom: distance of the selected source from the separatrix.}
\end{figure*}

Figure~\ref{fig:supp_spin_branch} follows the selected sourced track from
$a=0.99$ to $a=0.9999$ and inclinations $23^\circ$--$25^\circ$.  The track
remains within $1.3\times10^{-2}M$ of the inclined separatrix while preserving
stable access to the ZDM pole half-width.  It is therefore the numerical
continuation behind the near-extremal branch used in the Letter.

\section{Numerical accuracy check}

The pole tomography uses complex $\Zinf$ values from the high-resolution
$a=0.9999$, $e=0.1$, $(m,n,k)=(2,2,1)$ grid near the selected branch.  Table
\ref{tab:numerical_accuracy} reports the \texttt{pybhpt} radial-solver
precision diagnostics for the infinity and horizon amplitudes in this grid.
These diagnostics are several orders of magnitude below the order-unity
changes tested by the pole/non-pole tomography.

\begin{table}[t]
 \caption{\label{tab:numerical_accuracy}
 Numerical precision diagnostics for the high-resolution single-mode
 Teukolsky grid used in the pole tomography.  The entries give the median and
 maximum diagnostic values for the infinity and horizon amplitudes.}
 \begin{ruledtabular}
 \begin{tabular}{lccc}
 Sample & Channel & $N$ & median/max \\
 \hline
 finite grid &
 $\infty$ &
 262 &
 $1.9\times10^{-14}/5.6\times10^{-12}$ \\
 finite grid &
 $H$ &
 262 &
 $1.6\times10^{-14}/4.7\times10^{-12}$ \\
 $d_{\rm ZDM}\leq1$ &
 $\infty$ &
 129 &
 $2.7\times10^{-14}/5.6\times10^{-12}$ \\
 $d_{\rm ZDM}\leq1$ &
 $H$ &
 129 &
 $2.4\times10^{-14}/4.7\times10^{-12}$ \\
 $\iota=22^\circ$ cut &
 $\infty$ &
 16 &
 $1.5\times10^{-14}/8.1\times10^{-13}$ \\
 $\iota=22^\circ$ cut &
 $H$ &
 16 &
 $1.4\times10^{-14}/7.2\times10^{-13}$ \\
 \end{tabular}
 \end{ruledtabular}
\end{table}

The nearest sample to the benchmark crossing,
$(p,\iota)=(1.190,22^\circ)$, has
$d_{\rm ZDM}=4.06\times10^{-2}$,
$|\Zinf|^2=2.31\times10^5$,
$\epsilon_\infty=2.22\times10^{-13}$, and
$\epsilon_H=2.00\times10^{-13}$.  The same sample has negative horizon flux,
so it lies in the superradiant regime used in the main analysis.

\section{Pole tomography and pole fraction}

On every fixed-inclination cut we fit a quadratic smooth non-pole component
plus the independent ZDM pole and compare it with an equal-complexity cubic
polynomial.  The comparison uses leave-one-out complex-amplitude RMSE.  Varying
the fit window in $d_{\rm ZDM}$ and the non-pole order preserves the preference
for the fixed-pole model.  The fitted residue includes the Green-function
residue, source projection, and normalization conventions; consequently, the
pole fraction quoted below is a local response diagnostic, not an independently
normalized QNM excitation coefficient.

The joint tomography uses 110 samples on 12 inclination cuts, each containing
6--10 real-frequency points.  At the independent ZDM, the normalized joint
leave-one-out score is $0.0528972$.  The freely recovered pole has
$\Delta\omega_R/\gamma=-0.03125$, half-width ratio $1.02917$, and score
$0.0224826$.  Frequency shifts by one pole half-width, halving or doubling the
pole half-width, and reflection into the upper half plane all perform
substantially worse.  A leave-one-inclination-out jackknife returns the same
grid minimum for all 12 omissions.

\begin{figure*}[t]
 \includegraphics[width=\textwidth]{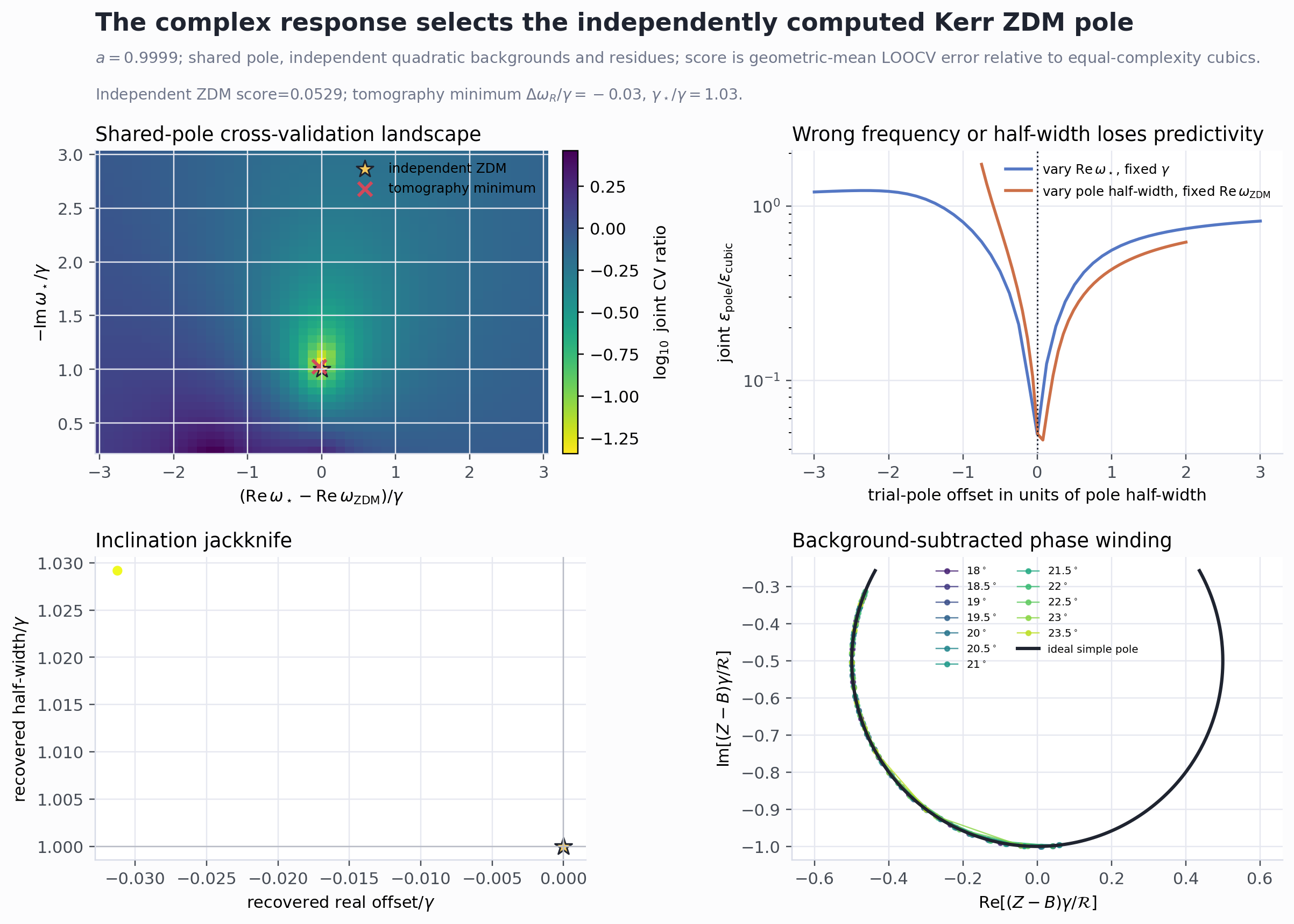}
 \caption{\label{fig:supp_tomography}
 Tomography robustness diagnostics.  Top left: joint leave-one-out
 cross-validation score in the complex trial-pole plane; the star marks the
 independently computed ZDM and the cross marks the best recovered pole.  Top
 right: one-dimensional coordinate slices showing that shifting the real
 frequency or changing the pole half-width worsens the prediction error.
 Bottom left: leave-one-inclination-out jackknife of the recovered pole
 coordinates, testing that no single inclination cut fixes the result.  Bottom
 right: residue-normalized, background-subtracted phase winding for each
 inclination cut compared with the ideal simple-pole trajectory.  Together the
 panels show that the complex response selects the independent Kerr ZDM pole.}
\end{figure*}

Figure~\ref{fig:supp_tomography} displays the pole-recovery controls.  The
cross-validation landscape and one-dimensional slices test the two coordinates
of the complex pole; the jackknife tests that no single inclination cut fixes
the result; and the phase-winding panel checks the residue-normalized
simple-pole behavior after smooth-background subtraction.

For each local fit we evaluate
\begin{equation}
 {\cal P}=\frac{|{\cal R}/(\omega-\omega_{\rm ZDM})|}{|B(\omega)|}.
\end{equation}
The full sample has median ${\cal P}=0.632195$, maximum $0.886629$, and no
sample with ${\cal P}>1$.  For $d_{\rm ZDM}<0.1$, the range is
$0.738$--$0.887$, and the rank correlation between ${\cal P}$ and
$d_{\rm ZDM}$ is $-0.923$.  The pole contribution is therefore comparable to,
but not larger than, the fitted smooth non-pole part in the local response
model.

\section{Eccentricity and inclination survey}

The fixed-spin survey at $a=0.9999$ spans
$0.01\leq e\leq0.50$ and $5^\circ\leq\iota\leq30^\circ$ for the selected
orbital harmonic.  It checks whether nearby eccentric and inclined geodesics
retain stable access to the same near-ZDM condition.  The scan is not a
universality statement over all EMRIs; it tests the local robustness of the
representative branch used in the Letter.

\begin{figure*}[t]
 \includegraphics[width=\textwidth]{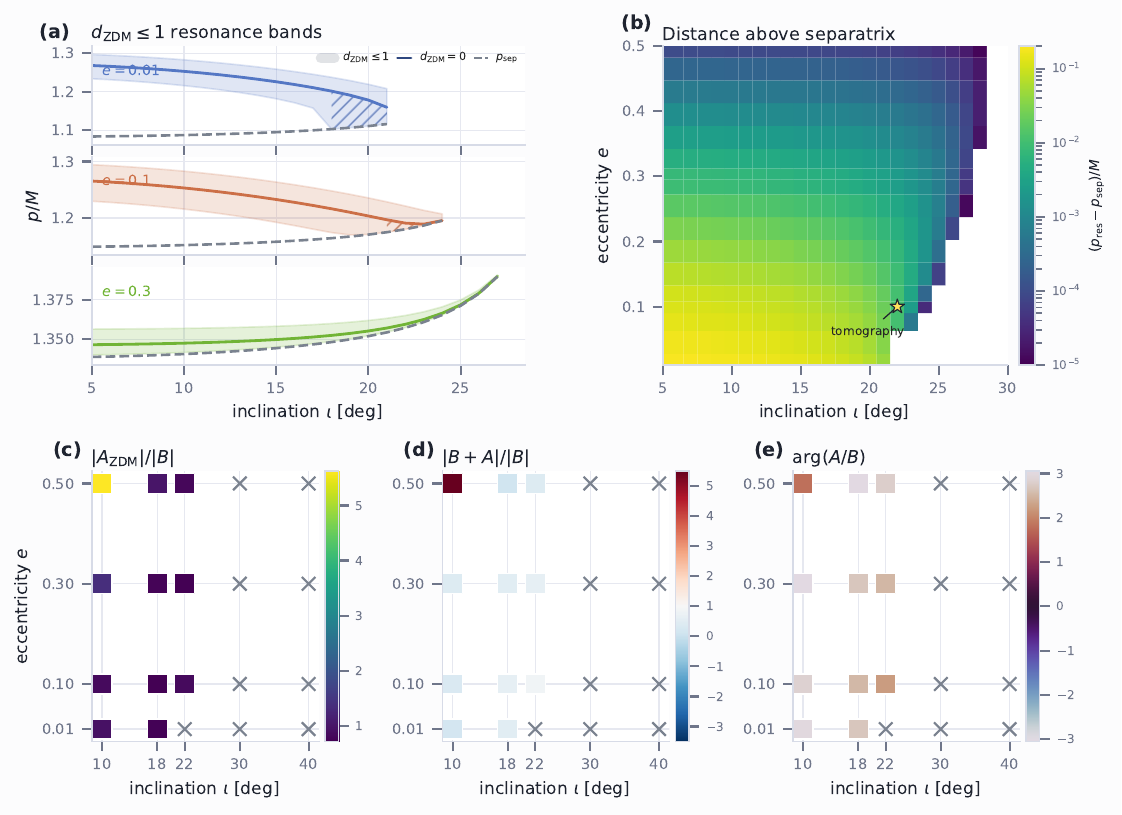}
 \caption{\label{fig:supp_parameter_survey}
 Eccentricity--inclination survey at $a=0.9999$.
 (a) Connected $d_{\rm ZDM}\leq1$ bands for representative eccentricities,
 with hatching marking separatrix-truncated bands.  (b) Stable resonance
 distance above the separatrix over the scan.  (c)--(e) Source projection,
 coherent gain, and pole--background phase at stable resonance roots; crosses
 mark samples without a usable stable root or local fit.}
\end{figure*}

Figure~\ref{fig:supp_parameter_survey} shows that the near-ZDM band persists
over nonzero inclinations and across the sampled eccentricities, while the
source strength and separatrix distance vary across the scan.

\end{document}